 \documentclass[final,5p,times,twocolumn]{elsarticle}
 
 \usepackage{epsfig}

\usepackage{amssymb}
 \usepackage{amsthm}
 
 \usepackage{graphicx}% Include figure files
\usepackage{dcolumn}% Align table columns on decimal point
\usepackage{bm}% bold math
\usepackage{subfig}%
\usepackage{caption}
\usepackage{hyperref}% add hypertext capabilities

\journal{NIM B}

\usepackage[switch,columnwise]{lineno}
\begin{document}

\begin{frontmatter}

%% Title, authors and addresses

%% use the tnoteref command within \title for footnotes;
%% use the tnotetext command for theassociated footnote;
%% use the fnref command within \author or \address for footnotes;
%% use the fntext command for theassociated footnote;
%% use the corref command within \author for corresponding author footnotes;
%% use the cortext command for theassociated footnote;
%% use the ead command for the email address,
%% and the form \ead[url] for the home page:
%% \title{Title\tnoteref{label1}}
%% \tnotetext[label1]{}
%% \author{Name\corref{cor1}\fnref{label2}}
%% \ead{email address}
%% \ead[url]{home page}
%% \fntext[label2]{}
%% \cortext[cor1]{}
%% \address{Address\fnref{label3}}
%% \fntext[label3]{}

\title{Measurement of the equilibrium charge state distributions of Ni, Co, and Cu beams in Mo at 2 MeV/u: review and evaluation of the relevant semi-empirical models.}

%% use optional labels to link authors explicitly to addresses:

 \author[label1,label2]{P. Gastis} 
\cortext[mycorrespondingauthor]{Corresponding authors}
\ead{gasti1p@cmich.edu}
 
 \author[label1,label2,label3]{G. Perdikakis}
 \ead{perdi1g@cmich.edu}

 \author[label2,label4]{D. Robertson}
 \author[label1]{R. Almus}
 \author[label2,label4]{T. Anderson}
 \author[label4]{W. Bauder}
 \author[label2,label4]{P. Collon}
 \author[label2,label4]{W. Lu}
 \author[label2,label4]{K. Ostdiek}
 \author[label2,label4]{M. Skulski}

 \address[label1]{Department of Physics, Central Michigan University, Mt. Pleasant. MI 48859, USA}
 \address[label2]{Joint Institute for Nuclear Astrophysics: CEE, Michigan State University, East Lansing, MI 48824, USA}
 \address[label3]{National Superconducting Cyclotron Laboratory, Michigan State University, East Lansing, MI 48824, USA}
 \address[label4]{Department of Physics, University of Notre Dame, Notre Dame, IN 46556, USA}

%\author{P. Gastis}
%\address{gasti1p@cmich.edu}

\begin{abstract}
Equilibrium charge state distributions of stable $^{60}$Ni, $^{59}$Co, and $^{63}$Cu beams passing through a 1$\mu$m thick Mo foil were measured at beam energies of 1.84 MeV/u, 2.09 MeV/u, and 2.11 MeV/u respectively. A 1-D position sensitive Parallel Grid Avalanche Counter detector (PGAC) was used at the exit of a spectrograph magnet, enabling us to measure the intensity of several charge states simultaneously. The number of charge states measured for each beam constituted more than 99\% of the total equilibrium charge state distribution for that element. Currently, little experimental data exists for equilibrium charge state distributions for heavy ions with 19$\lesssim Z_{p},Z_{t} \lesssim$54 (Z$_{p}$ and Z$_{t}$, are the projectile's and target's atomic numbers respectively). Hence the success of the semi-empirical models in predicting typical characteristics of equilibrium CSDs (mean charge states and distribution widths), has not been thoroughly tested at the energy region of interest. A number of semi-empirical models from the literature were evaluated in this study, regarding their ability to reproduce the characteristics of the measured charge state distributions. The evaluated models were selected from the literature based on whether they are suitable for the given range of atomic numbers and on their frequent use by the nuclear physics community. Finally, an attempt was made to combine model predictions for the mean charge state, the distribution width and the distribution shape, to come up with a more reliable model. We discuss this new "combinatorial" prescription and compare its results with our experimental data and with calculations using the other semi-empirical models studied in this work.
\end{abstract}

\begin{keyword}
%% keywords here, in the form: keyword \sep keyword
charge state distributions \sep semi-empirical models \sep heavy ion beams \sep  molybdenum foil \sep gas cell windows 
%% PACS codes here, in the form: \PACS code \sep code

%% MSC codes here, in the form: \MSC code \sep code
%% or \MSC[2008] code \sep code (2000 is the default)

\end{keyword}

\end{frontmatter}

%% \linenumbers

%% main text
\section{\label{sec:level1}Introduction}

The charge state distributions (CSDs) of heavy ions often have to be considered in accelerator design and in the development of stable and radioactive isotope spectrometers. An important need of information on charge state distributions arises with the use of recoil mass spectrometers (such as the electromagnetic mass analyser, EMMA \cite{emma} at TRIUMF, Canada, and the fragment mass analyser, FMA \citep{fma} at Argonne National Laboratory) to study nuclear reactions in inverse kinematics. These systems use electromagnetic fields to filter out the reaction recoils from the unreacted beam and rely critically on the fact that ions with different m/Q ratios (where m is the mass of the ion and Q its charge) follow in principle different trajectories inside a field. 
In experiments with recoil spectrometers the usage of gas-cell targets is common. As a beam of recoil products passes through the windows of a gas cell it interacts with electrons in the material via atomic charge-exchange processes (electron captures and losses). A CSD associated with the material and thickness of the window is observed for the particles exiting the target. The different charge states of the recoils follow different trajectories inside systems employing electric or magnetic fields. Due to the typically limited acceptance of systems such as magnetic dipoles, only a small number of charge-states (in some cases only one) can be transmitted through them and hit the detector for a given field setting. Hence, to determine reaction yields using a recoil spectrometer, one needs to know their charge state distribution.
Molybdenum (Mo) foils are extensively used as windows in gas-cells since they offer various advantages. The high resistance of Mo under mechanical stress (large Young's modulus \citep{gas_c}) allows the achievement of high gas pressures using thin windows. Furthermore, Mo may reduce significantly the background from fusion evaporation reactions compared to a lower-Z foil. 

In the present study we measured the CSDs of stable $^{60}$Ni, $^{59}$Co, and $^{63}$Cu beams while passing through 1$\mu$m Mo foils (Z$_{t}$=42). The beam energies were 1.84 MeV/u, 2.09 MeV/u, and 2.11 MeV/u for the Ni, Co, and Cu respectively. The results of this study were used to check the agreement of semi-empirical models developed for heavy ions to the experimental CSDs. In the following we present a brief description of the methods for the calculation of CSDs (subsection \ref{subsec:calc}), the formalism for the semi-empirical models considered (subsection \ref{subsec:Models}), the experimental procedure followed to extract CSDs at the University of Notre Dame (Section \ref{sec:Experiment}), our results in a detailed comparison and analysis of the successes and shortcomings of various semi-empirical models (Section \ref{sec:Resuls and Discussion}), and a conclusion and outlook (Section \ref{sec:Conclusion}).

\subsection{\label{subsec:calc} Methods for the calculation of charge state distributions}

The calculation of CSDs for systems with Z$>$2 is challenging since, for every ion/target combination of interest, a complete set of electron capture and loss cross sections must be known \citep{ben72}. 
Currently, a number of techniques for the calculation of cross sections (e.g continuum distorted wave approximation, plane-wave Bohr approximation) and the treatment of the electron exchange processes (e.g quasiground-state model, three charge-state model, etc.) are adopted by computer codes in order to numerically solve the problem. For example, the codes CHARGE and GLOBAL have been designed for the calculation of equilibrium and non-equilibrium CSDs, at energies above 100 MeV/u, for heavy projectiles with atomic number ($Z_{p}\geq30$) in solid targets \citep{sch98}. For intermediate energies above 10 MeV/u (and up to about 30 MeV/u), the program ETACHA \citep{roz96} calculates the evolution of charge state distributions for ions with up to 28 electrons; the program has also been checked at energies around 2 MeV/u \citep{ima05}. Finally, the code CSDsim has been developed at calculation of equilibrium and non-equilibrium CSDs for beams and recoils in gas targets \citep{zyl07}. Generally speaking, for all the codes mentioned above the ability of CSD prediction is limited; theoretical calculations on electron exchange cross sections don't have the required accuracy, so experimental values must be used instead. Calculated CSDs in regions (energies and atomic numbers) where cross section measurements have been done or where the adopted approximations are correct (e.g at high energies) are more likely to be accurate than outside of these regions.

Apart from computational methods such as those described above, semi-empirical models are widely used to calculate equilibrium CSDs for practical applications (e.g on LISE++ \cite{tar2004}). 
Important parameters of the distributions, such as the mean charge state and the distribution width, are given by empirical formulas based on experimental data. Following that, symmetric (Gaussian) or asymmetric functions are used to model the distribution and to calculate the equilibrium fractions. The main advantage of semi-empirical models is the simplicity and speed of the calculations for any system. However, the calculations are strictly limited to equilibrium CSDs and for systems (ion/target) for which experimental data exist.
In the energy region up to 10 MeV/u for projectile and target atomic number combinations 19$ \lesssim Z_{p}, Z_{t} \lesssim $54 the semi-empirical models are expected to more accurately describe the charge state distributions than the detailed codes described earlier (CHARGE, GLOBAL, ETACHA). However, their agreement with experiment has not been tested extensively since very few experimental data exist for that region.

\subsection{\label{subsec:Models}Semi-empirical models for equilibrium charge state distributions: mean charge state and distribution width}

So far there is no quantitative theory to calculate from first principles the mean charge state of heavy ions passing through solids. 
However, in the last decades, experimental data on equilibrium CSDs have been accumulated in databases, such as \citep{Nov15a}, informing the development of semi-empirical calculations.
Difficulties are also encountered regarding the distribution widths.
For high-Z ions, the electron capture and loss cross sections change significantly across shell closures giving rise to obvious asymmetries in the CSD data (shell effects) \citep{moak67,datz71}. For such CSDs the notion of a single distribution width \textit{d} is not valid any more. However, since an analytical calculation of CSD for a heavy ion is very challenging, single widths are still in use in order to approximately model charge distributions. It is common practice to calculate experimental widths assuming a Gaussian distribution (see  Eq. \ref{stand} below), even for those cases for which non Gaussian distributions are expected. Nevertheless, based on the existing experimental data, a number of such empirical models have been developed for the estimation of equilibrium distribution widths in gaseous and solid targets. 
%(\emph{-------Which models are these? Sciwietz?---}-). 

A number of semi-empirical models from the literature were considered in this study, based on their validity in the relevant atomic number, energy range, and on their frequent use by the nuclear physics community to predict characteristics of CSD.  In this section we will briefly present the basic formulation of each of these models and comment on their performance.

The most recent semi-empirical model has been developed by G. Schiwietz et al. \citep{sch04}. A many-parameter formula has been fitted on data for about 840 experimental CSDs at various energies and atomic number ranges for the projectile and target. More specifically the data includes CSDs from solid targets with 4$\leq $Z$_{t}$ $\leq$83 (although more than  40\% of the experimental distributions corresponds to C foils), and projectile ions with 1$\leq$ Z$_{p}$ $\leq$92, at energies up to $\sim$50 Mev/u. The formula extracted by the fitting procedure is:
\begin{eqnarray}
\overline{q}=\frac{Z_{p}(8.29x+x^{4})}{0.06/x+4+7.4x+x^{4}}
\label{schiw}
\end{eqnarray}  
where:
\begin{eqnarray}
x=c_{1}(\tilde{\upsilon}/c_{2}/1.54)^{1+1.83/Z_{p}},
\label{1}
\end{eqnarray} 
\begin{eqnarray}
c_{1}=1-0.26e^{-Z_{t}/11}e^{-\frac{(Z_{t}-Z_{p})^{2}}{9}},
\label{2}
\end{eqnarray} 
\begin{eqnarray}
c_{2}=1+0.03\tilde{\upsilon} ln(Z_{t}),
\label{3}
\end{eqnarray} 
with the so-called scaled velocity $\tilde{\upsilon}$, given in terms of the projectile's velocity $\upsilon _{p}$ and Bohr velocity $\upsilon _{B}$, by:
\begin{eqnarray}
\tilde{\upsilon}=Z_{p}^{-0.543}\upsilon _{p}/\upsilon _{B}.
\label{4}
\end{eqnarray} 
 According to their study, deviations from experiment on the mean charge-state values, $\overline{q}$, are of the order of 2\%. For the distribution widths, G. Schiwietz et al. in \citep{sch01} proposes a relation of the form:
\begin{eqnarray}
d = w [Z_{p}^{-0.27} Z_{t}^{0.035-0.0009 Z_{p}} f(\overline{q})f(Z_{p}-\overline{q})]^{-1}
\label{sc}
\end{eqnarray}
where \textit{w} is a scaled width and 
\begin{eqnarray}
f(x)=\sqrt{(x+0.37Z_{p}^{0.6})/x}.
\end{eqnarray}
 %(-----the following phrase is somehow incomprehesible for me Can you explain?-----why is w$\approx$0.7 a satisfactory value...).
In \citep{sch01}, the width \textit{w} is plotted versus the number of bound electrons, N$_{b}$, of the projectile where N$_{b}$=Z$_{p}$ - $\overline{q}_{exp.}$. According to the scaled solid-state data, \textit{w}$\approx$0.7, which is satisfactorily accurate in our present data (N$_{b} \approx$10). 
 
In a different approach, (but also based on available experimental data) J.A Winger et al. \citep{winger} developed a phenomenological parameterised formula for the mean charge state which has the form:
 \begin{eqnarray}
\overline{q}=Z_{p}[1 - exp(\sum_{i=0}^{4} \alpha _{i} X^{i})]
\label{wing}
\end{eqnarray}  
where the reduced velocity X is given in terms of the beam's kinematic factor $\beta$, as:
\begin{eqnarray}
X=\beta /0.012Z_{p}^{0.45}
\label{red}
\end{eqnarray}  
The parameters $\alpha _{i}$ are defined analytically in the way presented in \citep{winger}. No additional information is given about the model's performance and agreement with experiment. In the same study, a phenomenological parametrization for the widths is also derived. According to this model:
\begin{eqnarray}
d = exp(\sum _{i=0}^{i=2} \alpha _{i} (lnX)^{i})[1-exp(\sum _{i=0}^{i=2} \beta _{i} (ln Z_{t})^{i})]
\label{w}
\end{eqnarray}
The parameters $\alpha _{i}$ and $\beta _{i}$ are different in this formula from those in the previous equation (for details see ref. \cite{winger}). In order to avoid large deviations due to shell effects they used experimental data at energies above $\sim$5 MeV/u. The energy of ions in the present study is outside this limit and hence, widths calculated with Winger's  formula are expected to show worse agreement with experiment than some of the other more suitable models we considered.

In an older study, Nikolaev and Dmitriev (ND) \cite{nik} derived a semi-empirical formula for the mean charge states, according to which:
\begin{eqnarray}
\overline{q}=Z_{p}[1+(\upsilon/Z_{p}^{\alpha}\upsilon ')^{-1/k}]^{-k}
\label{nd}
\end{eqnarray}
where $\alpha$=0.45, \textit{k}=0.6, $\upsilon$ is the projectile's velocity, and $\upsilon$'=3.6x10$^{8}$cm/sec. Eq. (\ref{nd}) is designed to reproduce data with Z$_{p} \gtrsim$ 20, at energies of 5 to 200 MeV, mainly on C targets. Deviations from experiment on the $\overline{q}$ values do not exceed 5\%, according to \citep{nik}. In the same study, an improved formula for the distribution widths optimized for solid ion beam strippers is also presented:
\begin{eqnarray}
d = d_{0}[\overline{q}[1-(\overline{q}/Z_{p})^{1/k}]]^{1/2}
\label{nik}
\end{eqnarray}
where d$_{0}$=0.5 and k=0.6. Eq. (\ref{nik}) is expected to be more accurate for Z$_{p}\lesssim$37 at energies above 20 MeV, as discussed in \citep{ben72}. Based on the ND model, E. Baron et al. \citep{baron} developed their own empirical formula to predict the average charge states $\overline{q}$, for ion species in the range 18$\leq Z_{p} \leq$92 and energies up to 10.6 MeV/u. According to their improved model:
\begin{eqnarray}
\overline{q}=Z_{p}[1-Cexp(-83.275 \beta / Z_{p}^{0.477}]\nonumber\\
\times
[1-exp(-12.905+0.2124Z_{p}-0.00122Z_{p}^{2})].
\label{bar}
\end{eqnarray}
where C=1 for energies E$_{p}>$1.3~MeV/u and C=0.9+0.0769E$_{p}$, for E$_p<$ 1.3~MeV/u. %(+++++Accuracy expectation accordign to authors?++++).
Their empirical formula for the widths has the form: 
\begin{eqnarray}
d = \sqrt{\overline{q}(0.07535+0.19Y-0.2654Y^{2})}
\label{bb}
\end{eqnarray}
where Y=$\overline{q}$/Z$_{p}$. The width model was designed to be more accurate for Z$_{p}>$54 at energies above 1.3 MeV/u. 

In addition to the above works for the mean charge state and width of the charge state distribution, in the present study we also considered four more empirical models that provided formulas either for the mean charge state only, or for the distribution width. In particular these are: the work of H. D. Benz \cite{ben72} for the distribution width, and the works of K. Shima et al. \cite{shima82}, To and Drouin \cite{drouin}, and A. Leon et al. \cite{leon} for the mean charge state. These models are presented in detail below; H. D. Benz, based on Nikolaev-Dmitriev's work \citep{nik}, provides the following simple relation for the distribution width \textit{d}:
 \begin{eqnarray}
d=0.27Z_{p}^{1/2}
\label{benz}
\end{eqnarray}
which provides a fair agreement with experimental data for heavy ions, up to Uranium, in Carbon and Formvar foils, at energies below 80 MeV \citep{ben72}. K. Shima et al. \cite{shima82} developed a model for the mean charge states for a large range of ion species (Z$_{p}\geq$8) in solid targets with 4$\leq Z_{t} \leq$79 at energies below 6 MeV/u. According to his study:
\begin{eqnarray}
\overline{q}=Z_{p}[1-exp(-1.25X+0.32X^{2}-0.11X^{3})]\nonumber\\
\times
[1-0.0019(Z_{t}-6)\sqrt{X}+0.00001(Z_{t}-6)^{2}X],
\label{shi}
\end{eqnarray}  
where the scaled velocity X is given by Eq. (\ref{red}). The second part in Eq. \ref{shi} is actually a correction term for the non carbon solid targets. This model reproduces the experimental data of Z$_{p} \geq$14 with an agreement of $\Delta \overline{q} /Z_{p} <$0.04. A reformulation of the ND relation \citep{nik} is given by the model of To and Drouin \cite{drouin}, aiming to reproduce lighter elements (B to Ne) at energies up to 7 MeV on C targets. In this model the average charge state formula is expressed as:
\begin{eqnarray}
\overline{q}=Z_{p}[1-exp(-\upsilon /\upsilon ' Z_{p}^{0.45})].
\label{to}
\end{eqnarray}
Even though the formulas in the work of To and Drouin were not formulated to reproduce heavy ion/target combinations, their work was included in the current study since it is a derivative of the generally successful Nikolaev and Dmitriev model. Finally, A. Leon et al. \cite{leon}, based on Baron's work \citep{baron}, reformulated the mean charge state expression by multiplying Eq. (\ref{bar}) with a suitable correction factor g'(Z$_{t}$,Z$_{p}$).  This led to improved fits of experimental data at energies of 18 MeV/u $\leq E_{p} \leq$ 44 MeV/u for heavy ions with 36$\leq Z_{p} \leq$92 in various solid targets (4$\leq Z_{t} \leq$79).The multiplicative correction factor is given by:
\begin{eqnarray}
g\prime (Z_t,Z_p)=[(0.929+0.269 exp(-0.160 Z_{t}))+\nonumber\\
+(0.022-0.249 exp(-0.322Z_{t}))\frac{v_{p}}{Z_{p}^{0.477}}]
\label{leo}
\end{eqnarray}
where $\upsilon_{p}$ is the projectile's velocity.

\subsection{Modelling of the equilibrium charge-state distributions}

To succesfully calculate charge state fractions, apart from the mean charge state and the width, a model of the shape of the distribution is also needed.
At low and intermediate projectile velocities in light gaseous and solid targets (Z$_{t} \lesssim$7), the equilibrium CSDs tend to be symmetrical. For those cases, the charge fractions can be calculated using Gaussian functions:
\begin{eqnarray}
\label{gau}
F_{q}=(d\sqrt{2\pi})^{-1} \,\, exp[-(q-\overline{q})^{2}/2d^2]
\end{eqnarray}
where \textit{d} is the distribution width, q is the charge of each state (an integer number), and $\bar{q}$ is the mean charge state of the distribution (in general a real number). For higher projectile velocities in the same targets, typically for cases in which the mean charge state is very close to the Z$_{p}$, the CSDs become asymmetrical (these asymmetries are explicitly dependent on the velocity and are not related with the shell effects that we will discuss next). For those distributions, Baudinet-Robinet et al. \citep{rob} proposed a distribution function extracted from a reduced $\chi^{2}$ distribution:
\begin{eqnarray}
\label{ro}
F_{t}=[2^{\nu /2}\Gamma (\nu /2)]^{-1} t^{\nu /2-1} e^{-t/2}
\end{eqnarray}
where the chi-squared variable is connected to the charge q, mean charge $\overline{q}$, and width \textit{d} as: t=c(Z$_{p}$+2-q), c=2(Z$_{p}$+2-$\overline{q}$)/d$^{2}$ and $\nu$=c(Z$_{p}$+2 - $\overline{q}$). 

The CSDs of heavy projectiles, especially in heavy gases and solids are asymmetric mainly due to atomic shell structure. Those cases are far from Gaussian or $\chi^{2}$ distributions. K. Shima et al.\citep{shima83}, proposed the composite of two Gaussian functions with the same centroid but different standard deviations (widths) to emulate the asymetrical distributions of experiment. The now different left and right widths are associated with the different atomic shells. The usage of such a function in the calculation of CSDs requires a model for the estimation of double widths which hasn't been developed so far. 
Another approach for treating the shell effects has been proposed by R. O. Sayer \cite{sayer} who introduced a modified Gaussian distribution:
\begin{eqnarray}
\label{say}
F_{q}=F_{m}\, exp[-0.5t^{2}/(1+\epsilon t)]
\end{eqnarray}
where t=(q - q$_{0}$)/$\rho$ and F$_{m}$ is the fraction of the most intense charge state q$_{0}$. The shell effects can be satisfactorily reproduced (even if they are not explicitly taken into account in Eq. \ref{say}) if the proper values for $\rho$ and $\epsilon$ are chosen. These values can be extracted by fitting a suitable function of Z$_{p}$ and projectile velocity, $\beta$c, on experimental data. Since there is not a satisfactory amount of data in the  heavy ions region (Z$_{p}$, Z$_{t}>$20) the agreement of this method to experimental CSDs may be somewhat limited. 
%(+++++can we provide these widths from our own data and the library?+++++. 
Furthermore, the F$_{m}$ values must been known in advance.
Because of these limitations on Sayer's and Shima's formulas, Eq. (\ref{gau}) and Eq. (\ref{ro}) are generally preferred by the community for the calculation of CSDs (within their application  limits) since they can be directly be combined with a large variety of semi-empirical formulas for the mean charge state $\overline{q}$ and the distribution width \textit{d}. In this work we followed the same logic and decided to leave out the works of Sayer and Shima \citep{sayer},\citep{shima83} from our comparison to experiment. Hence, for the modeling of the charge-state distributions we used only the  Gaussian and the Baudinet-Robinet reduced $\chi^{2}$ distribution shapes described above.

%\newpage 
\section{\label{sec:Experiment}Experimental Procedure and Data Analysis}

\begin{figure}[]
\begin{center}
\includegraphics[width=0.85\linewidth]{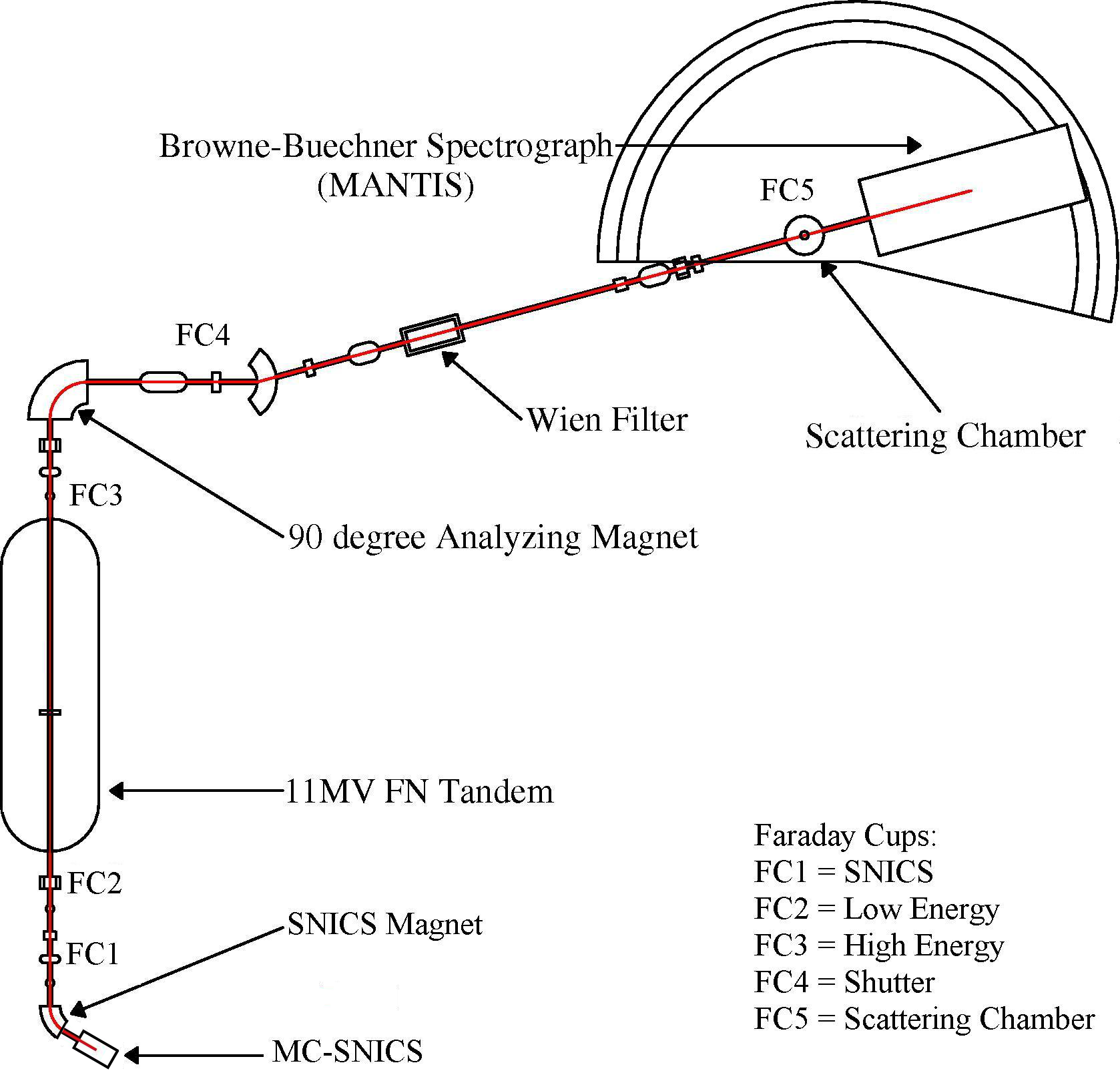}
\caption{\label{fig:wide}Schematic overview of the FN Tandem accelerator and the Accelarator Mass Spectroscopy beamline at the Nuclear Science Laboratory of the University of Notre Dame. The measurements in this work made use of the MANTIS Browne-Buechner Spectrograph, the Multi-purpose Rotational Scattering Chamber and the FN Tandem. The Wien filter shown in this figure upstream of the scattering chamber was not required and was not used in the experiment.}
\end{center}
\end{figure}

The experiment took place at the Nuclear Science Laboratory (NSL) of the University of Notre Dame. The incident ion beams were accelerated by the 11MV FN Tandem Van de Graaff accelerator. The accelerator mass spectrometry (AMS) beamline guided them into the Multi-purpose Rotational Scattering Chamber at the object of the MANTIS spectrograph (Fig. \ref{fig:wide}). For our measurements the spectrograph was set at a 0$^{\circ}$ angle with respect to the beam axis.  
%%%%%%%%%%%%%%%%%%%%%%%%%%%%%%%%%%%%%%%%%%%%%%%%%%%%%%%%%%%%%
%%%%%%%%%%%%%%%%%%%%%%%%%%%%%%%%%%%%%%%%%%%%%%%%%%%%%%%%%%%%%
Inside the scattering chamber a Faraday cup and three 1 $\mu$m thick Mo foil targets, were mounted on a movable metallic frame with five target positions. The fifth position was left blank so that the beam could pass through the chamber without interacting with the Mo foil. This setting was used during beam tuning. After the target ladder, the beam ions could enter the spectrograph where the different charge states of the beam could be separated by the magnetic field.

\begin{figure}[]
\begin{center}
\includegraphics[scale=0.16]{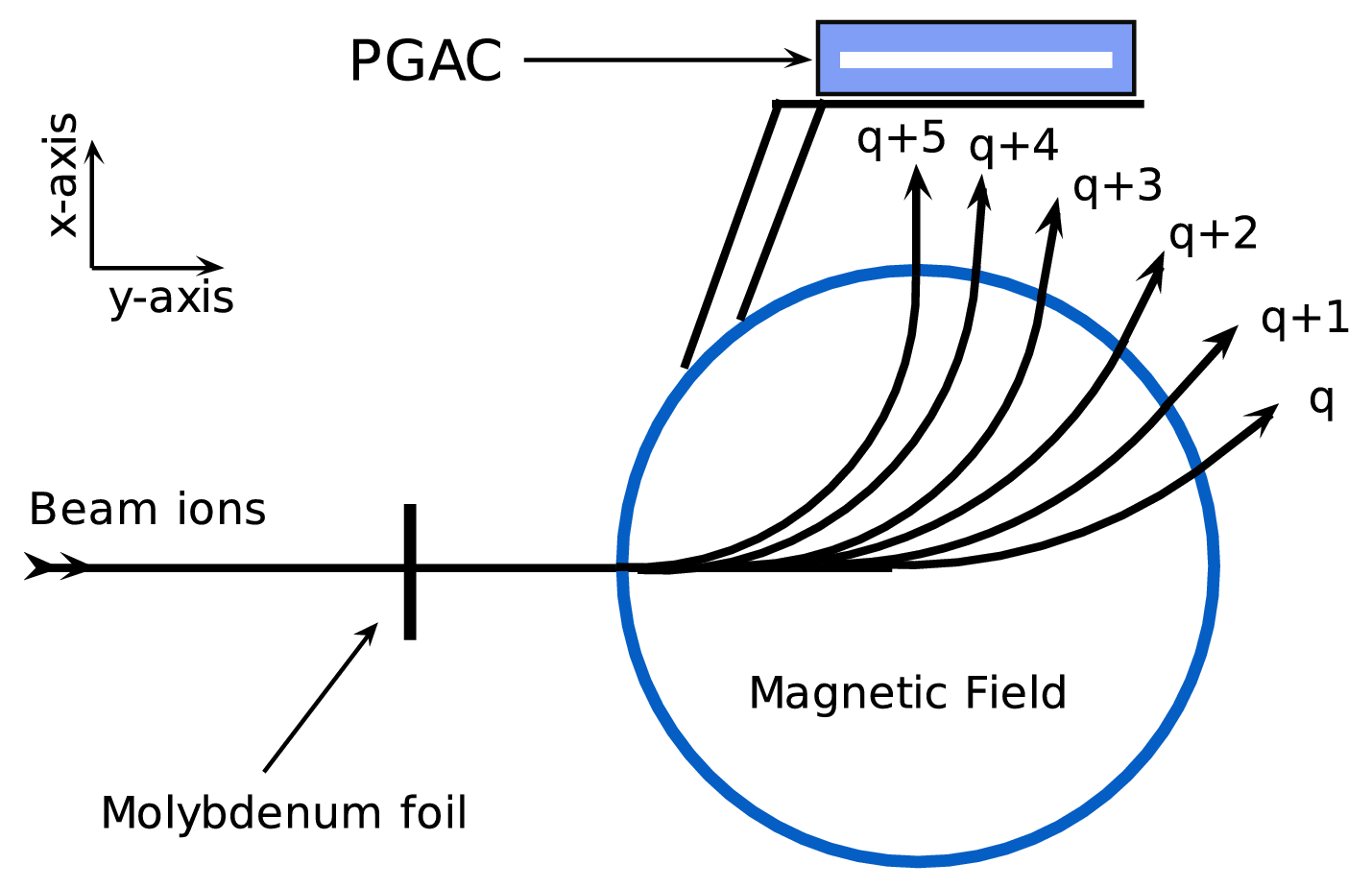}
\caption{\label{fig:wide2}Trajectories of the different charge states of the same element in the spectrograph magnet. The PGAC detector is position sensitive along the y-axis. The charge states are separated from each other by their charge-dependent position on the PGAC detector. The limit in the detector's active region allows the measurement of only those charge states that have a suitable trajectory radius in the magnet.}
\end{center}
\end{figure}

A Parallel Grid Avalanche Counter (PGAC) detector, with active region of 46 x 10 cm, was mounted on a set of rails on the top of the spectrograph magnet. At this point the bent beam leaves the magnet vertically as it is shown in Fig. \ref{fig:wide2}. By using the PGAC it was possible to measure a number of charge states simultaneously since the detector is position sensitive. A more detailed description of the detector's operation can be found in previous studies of D. Robertson et al. \cite{rob08, rob07}.
 
%%%%%%%%%%%%%%%%%%%%%%%%%%%%%%%%%%%%%%%%%%%%%%%%%%%%%%%%%%%%%
%%%%%%%%%%%%%%%%%%%%%%%%%%%%%%%%%%%%%%%%%%%%%%%%%%%%%%%%%%%%%

\begin{figure}[]
\begin{center}
\includegraphics[scale=0.39]{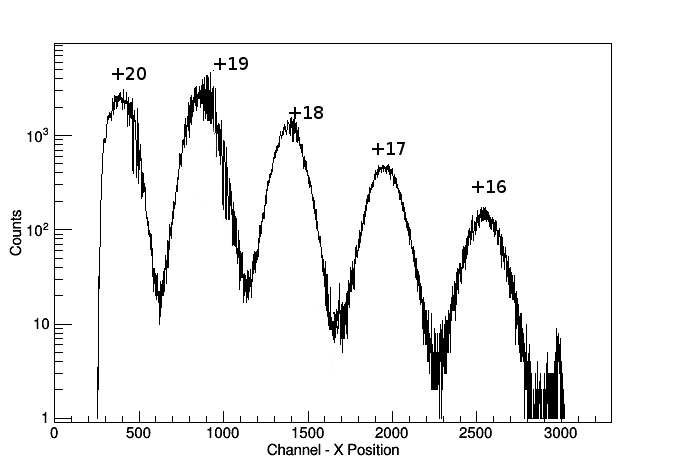}
\caption{\label{fig:wide3}Charge-states of Cu taken with the $^{63}$Cu beam. The spectrum corresponds to a single magnetic field setting. The channel number represents position along y-axis in the PGAC detector (0 cm to 46 cm). The intensity of each charge state is proportional to each peak's area.}
\end{center}
\end{figure}

The full charge-state distribution for each beam would not fit inside the magnetic spectrograph's acceptance in a single magnetic field setting. Therefore, a gradual step-by-step increase of the applied magnetic field was used to scan all detectable charge states using the detector's active region. In this process, by changing the field from lower to higher values the charge-states would be registered by the PGAC detector from higher to lower charge, since:
\begin{eqnarray}
\vert\vec{B} \vert \varpropto I_{magnet}\varpropto \frac{1}{q} 
\end{eqnarray}
where $\vec{B}$ is the magnetic field, $I_{magnet}$ is the current supplied to the spectrograph magnet, and a given trajectory radius through the magnet is assumed. Its important to mention that no focusing elements were used before or after the dipole spectrograph; so all charge states were transported to the detector with the same focusing characteristics as the beam along the dispersive direction of the magnet. The detector's length allowed the measurement of only 4 to 6 charge states at each step. As can be seen in Fig. \ref{fig:wide3} the charge states appeared in the spectra as different peaks along the horizontal axis which represents the Y-position inside the PGAC. Scattering in the Mo foil is a contributing factor in the observed widths of the peaks. 

 As mentioned before, because of the large number of charge states in each distribution, the measurements were performed in steps. Between two consecutive steps some charge states were chosen to be used as common references allowing the cross-normalization of the intensities of all charge states along the various spectra. 
%%%%%%%%%%%%%%%%%%%%%%%%%%%%%%%%%%%%%%%%%%%%%%%%%%%%%%%%%%%%%%%%%%%%%%%%
%%%%%%%%%%%%%%%%%%%%%%%%%%%%%%%%%%%%%%%%%%%%%%%%%%%%%%%%%%%%%%%%%%%%%
To deduce the charge-state distribution from the data the following procedure was used: For each charge-state q we defined the relative fraction R$_{q}$, calculated with respect to a reference state q$_{ref.}$:
\begin{eqnarray}
R_{q}=\frac{I_{q}}{I_{q_{ref}}}
\end{eqnarray}
where I$_{q}$ is the intensity of the state q and I$_{q_{ref.}}$ is the intensity of the reference charge state. By using the common charge states between adjacent steps as references we were able to calculate (for each distribution) all the relative fractions with respect to a single state. Having normalized the relative fractions in this way, the net fractions F$_{q}$ could be extracted by:
\begin{eqnarray}
F_{q}=\frac{R_{q}}{\sum_{q'} R_{q'}}.
\end{eqnarray}
where the sum is over the total number of charge states in the distribution. The uncertainties of the fractions were calculated from the statistical errors in peak integration and taking into account any overlap of adjacent peaks.

Having deduced the fractions of all the charge states, the mean charge of each distribution was calculated by: %$\overline{q}=\sum_{i=1}^{i=n} q_i F(q_i)$ 
\begin{eqnarray}
 \overline{q}=\sum_{q} q F_{q}
 \end{eqnarray}
and the distribution widths (assuming Gaussian distributions) by: %$d=\sum_{i=1}^{i=n} (q_i - \overline{q})^2 $ F(q_i)  
 \begin{eqnarray}
 d=[\sum_{q} (q - \overline{q})^2  F_{q}]^{1/2}
 \label{stand}
 \end{eqnarray}
%%%%%%%%%%%%%%%%%%%%%%%%%%%%%%%%%%%%%%%%%%%%%%%%%%%%%%%%

For each distribution measured in this experiment the registered charge states did not constitute 100\% of the whole CSD. The very low intensity charge states fell below the minimum detection limit of the experimental set-up. Moving toward the higher charge states the intensity drops very fast since the electron loss cross section changes significantly; due to the finite dispersion of the magnet the high charge states (low m/Q ratio) are expected to be closer to each other, and so, significantly more overlapped. These factors limited our ability to measure higher charge states in a reasonable time. Regarding the lower charge states, no reliable data could be taken for charge states with F$_q <10^{-1}$\% due to the post-foil interactions of the beam ions with the residual gas in the beamline vacuum. In such interactions it is expect that the electron capture cross sections are higher than the electron loss since the mean charge states after the Mo foil will tend to decrease in the residual air. Even by assuming -according to our calculations for the experimental beamline used- that only 0.1\% of the beam ions will interact with the residual gas, the effect on the intensity of the low charge states with F$_q <10^{-1}$\% is still significant due to the electron captures on higher charge states (especially on those with F$_q > 5$\%). These effects are hard to be estimated without further measurements or knowledge of the relevant cross sections.

The significance of the systematic uncertainty induced to the charge state fractions due to the missed charge states was estimated through a sensitivity test. Three more charge-states were added in each CSD and the effects on the charge fractions were calculated. The additional charge states had fractions F$_q = 10^{-1}$\% in order to maximize the effects. The charge state fractions varied as a result of this procedure by a factor which was found to be at most 0.09\%. Statistical errors coming from other contributing factors such as peak integration, fluctuated within a range of 1\% to 41\%. As a result, the error due to the missed charge states was considered negligible.

During the measurements no m/Q interferences from secondary particles (reaction products) were present. All the nuclear reaction channels from the interaction of the beams we used (Ni, Co, and Cu) with the Mo were found to have thresholds above 132 MeV while the beam energies we used were up to 125 MeV. Furthermore, contaminations in the beam from nuclear reactions with the carbon foil in the accelerator's terminal were eliminated by the 90 degree beam analysis magnet. Uncertainties due to possible pile-up were also negligible. Throughout the measurements the count rate in the detector was monitored and was found to be in no case more than of the order of 5000 counts/sec. These rates are too small to induce pile-up since the signal processing time in the counter detector used is of the order of a few microseconds for each particle.

\section{\label{sec:Resuls and Discussion}Results and Discussion}

\begin{figure}
\begin{center}
\includegraphics[scale=0.39]{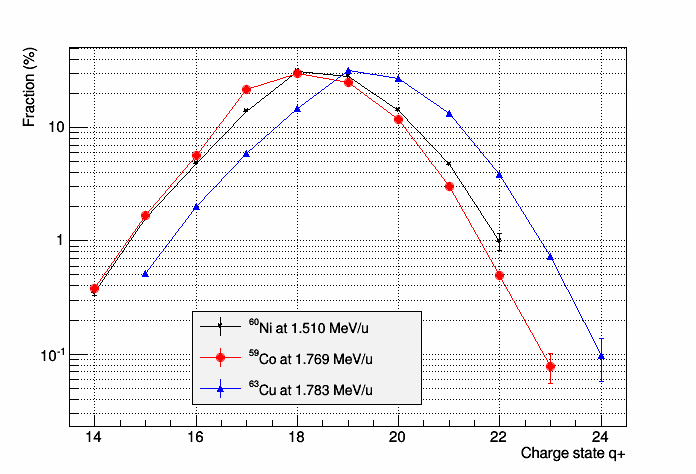}
\caption{\label{fig:csd}Charge state distributions of $^{60}$Ni (black inverted triangles and line), $^{59}$Co (red circles and line), and $^{63}$Cu (blue upright triangle and line) beams in a 1$\mu$m thick Mo foil. No error bars are visible for statistical errors less than 3\% due to the size of the point markers. The experimentally deduced mean charge states for each distribution are $\overline{q}$=18.45 for Ni, $\overline{q}$=18.20 for Co, and $\overline{q}$=19.33 for Cu. The most intense charge states were 18+, 18+, and 19+ for Ni, Co, and Cu respectively.}
\end{center}
\end{figure}

In Fig. \ref{fig:csd} the CSDs measured in the current study are presented. Furthermore, Table \ref{tab:tableA} includes the fractions of all the measured charge states in detail. The statistical error on the charge state fractions fluctuated between 1\% and 6\% in most of the charge states. The largest uncertainties were observed for the 22+ ($\sim$ 17\%), 23+ ($\sim$ 30\%), and 24+ ($\sim$ 41\%) charge states of Ni, Co, and Cu respectively. The increased errors at the highest charge states are due to the low statistics of the corresponding peaks in combination with the overlap of adjacent peaks for these higher charge states.
In order to be consistent with literature data, we assigned the CSDs to the emerging energies -i.e. for ions that have already lost energy by traversing the thickness of the target. In this way there is a direct correspondence between equilibrium CSD and projectile velocity (which is not the case when the initial beam energy is assigned). The energy loss was calculated using SRIM-2013 \citep{zie2010} and was found to be approximately 20 MeV for all three beams. The errors introduced to the mean charge state due to the energy loss calculation with SRIM were estimated to be less than 0.01 charge units.

\begin{figure}
\begin{center}
\includegraphics[scale=0.39]{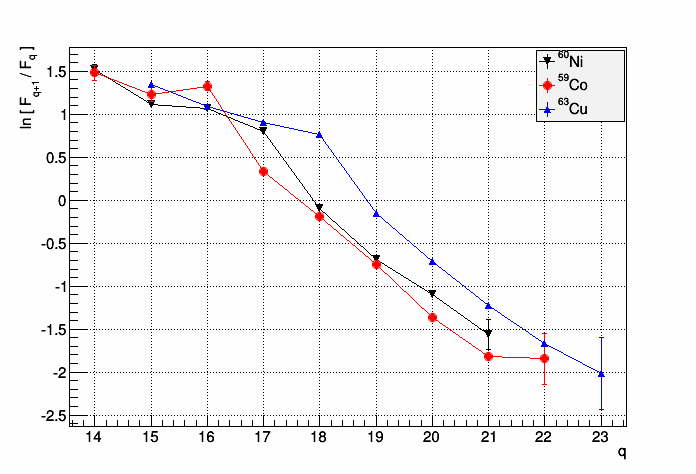}
\caption{\label{fig:cs}Logarithm of the ratio of two adjacent charge state fractions as function of the charge of ions detected with the MANTIS spectrograph in the present study. Lines are used to guide the eye. Black inverted traingles and a black line correspond to Ni ions on Mo using the $^{60}$Ni beam, red circles and line correspond to Co ions on Mo using the $^{59}$Co beam, and blue upright triangles and line correspond to Cu ions on Mo using the $^{63}$Co beam. Divergence from a straight line is attributed to shell effects and multiple-electron exchange processes. All observed distributions demonstrated such deviations which correspond to non-symmetrical CSDs.}
\end{center}
\end{figure}

All three distributions were found to be asymmetric due to shell effects. This asymmetry is evident by examining the ratios of the fractions for each CSD. In the case of a symmetric (Gaussian-like) equilibrium CSD, the logarithm of the ratios F$_{q+1}$/F$_{q}$ is linearly varying with the charge q. This is because, at equilibrium, the ratios $\sigma_{q,q+1}$/$\sigma_{q+1,q}$ (where $\sigma_{q,q+1}$ is the electron loss cross-section at the charge-state q and $\sigma_{q+1,q}$ is the electron capture cross-section at q+1) are approximately proportional to e$^{-q}$ and F$_{q+1}$/F$_{q}$=$\sigma_{q,q+1}$/$\sigma_{q+1,q}$ \cite{niko}. The above statement implies that the single electron exchanges are dominant. However, when shell transitions or multiple-electron exchanges occur the relationship between ln(F$_{q+1}$/F$_{q}$) and \textit{q} is not linear, resulting in asymmetrical CSDs. In Fig. \ref{fig:cs}, the discontinuities at q=16+, q=17+, and q=18+ (in Co, Ni, and Cu respectively) are consistent with the L-M shell transitions as can be shown by examining the corresponding electronic configurations \citep{NIST}. The fractions of charge states that correspond to closed shells, or sub-shells, are significantly enhanced resulting in "kinks" on the plot. For the case of Co, the discontinuity at q=22+ seems to correspond to the 2\textit{s}-2\textit{p} sub-shell transition. 
    
%%%%%%%%%%%%%%%%%%%%%%%%%%%%%%%%%%%%%%%%%%%%%%%%%%%%%%%%%%%%%%%%%%%%%%%%
%%%%%%%%%%%%%%%%%%%%%%%%%%%%%%%%%%%%%%%%%%%%%%%%%%%%%%%%%%%%%%%%%%%%%
%%%%%%%%%%%%%%%%%%%%%%%%%%%%%%%%%%%%%%%%%%%%%%%%%%%%%%%%%%%%%%%%%%%%%
In Table \ref{tab:table1}, all the calculations of the mean charge $\overline{q}$ using various semi-empirical models are presented. The experimental values have been calculated by Eq. (4). It is important to note that the mean charge of the distribution doesn't have to be an integer number. The maximum charge state (the most intense) is the integer which is closer to the mean value. In terms of their predictive power for the maximum charge state,  Schiwietz et al. and Shima et al. offer the best agreement to experimental data. The agreement of the other models examined in this work is within $\pm$2 charge units. The model by To and Drouin provided a better than expected agreement with experiment considering that it is designed to reproduce CSDs of lighter elements. On the other hand, the formula of Leon et al. (devised for heavier elements and higher energies) had the worst agreement of all models to our measurements. Regarding the mean value, Winger's model is in agreement to the experimental values within the uncertainty limits in most cases.

In Table \ref{tab:table2} the predictions of the semi-empirical models for the distribution widths are presented. The experimental widths have been calculated using Eq.(5). To compare each width formulation's performance to the others, the experimentally determined mean charge states $\overline{q}$ were used in all the calculations.
By neglecting shell effects, most of the models diverged from the experimental widths. A systematic overestimation was observed on the values calculated with Nikolaev-Dmitriev (ND) and Betz models, while the Baron et al. and Winger et al. models systematically underestimated the distribution widths. The model by Schiwietz et al. was shown to be in better agreement with experiment in reproducing the widths of the measured CSDs with very small deviations. This success we attribute to the fact that this model was based on fitted data that included a significant number of asymmetrical CSDs.  
%%%%%%%%%%%%%%%%%%%%%%%%%%%%%%%%%%%%%%%%%%%%%%%%%%%%%%%%%%%%%%%%%%%%%

%%%%%%%%%%%%%%%%%%%%%%%%%%%%%%%%%%%%%%%%%%%%%%%%%%%%%%%%%%%%%%%%%%%%%%%%

\begin{figure}[]
\begin{center}
\includegraphics[scale=0.39]{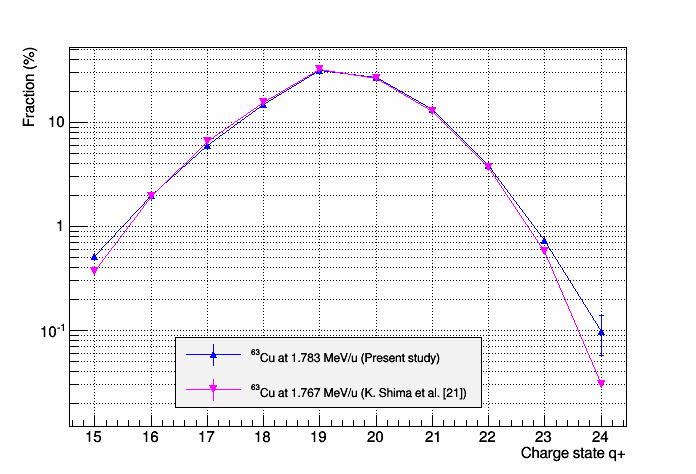}
\caption{\label{fig:liter}Equilibrium CSD of Cu in Mo, as measured in the present study (blue line, upright triangles) and as reported by K Shima et al \cite{shima86} (magenta line inverted triangles). The distributions are assigned to emerging projectile energies (after energy loss in the Mo foil). Mostly excellent agreement between the 2 measurements is observed.}
\end{center}
\end{figure}

The only available data in literature, relevant to the current work, are coming from a previous study by K. Shima et al. \citep{shima86}. Shima et al. used a $^{63}$Cu beam impinging on a Mo foil. The results are plotted in Fig. \ref{fig:liter} in comparison to data from the present work.  
Comparing the two data sets we see a mostly excellent agreement to each other considering the small energy difference of the two cases. The fractions of the charge states 19+ and 20+ are identical within the statistical errors while deviations are observed on charge-states with fractions lower than 15\%. In terms of the mean charge state and the distribution width (see Tables \ref{tab:table1} and \ref{tab:table2}), the deviations are within the statistical errors.

%%%%%%%%%%%%%%%%%%%%%%%%%%%%%%%%%%%%%%%%%%%%%%%%%%%%%%%%%%%%%%%%%%%%%%%%

%%%%%%%%%%%%%%%%%%%%%%%%%%%%%%%%%%%%%%%%%%%%%%%%%%%%%%%%%%%%%%%%%%%%%
%%%%%%%%%%%%%%%%%%%%%%%%%%%%%%%%%%%%%%%%%%%%%%%%%%%%%%%%%%%%%%%%%%%%%

\subsection{A "combinatorial" prescription for reproduction of the charge state distributions}
\begin{figure*}[]
\centering
\subfloat[]{\includegraphics[width=0.49\textwidth]{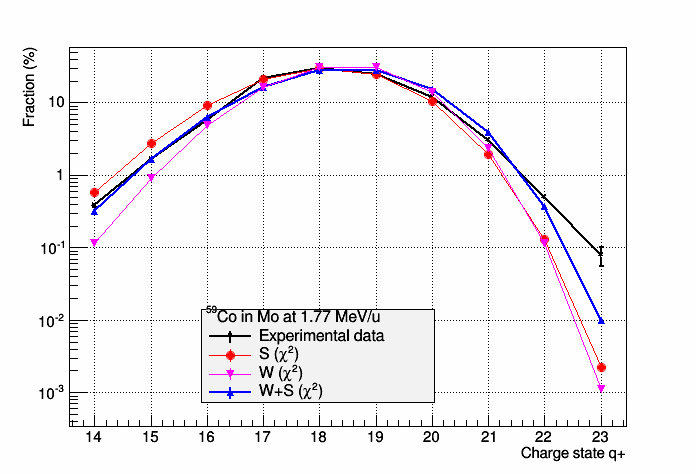}\label{fig:f1}}     
\subfloat[]{\includegraphics[width=0.49\textwidth]{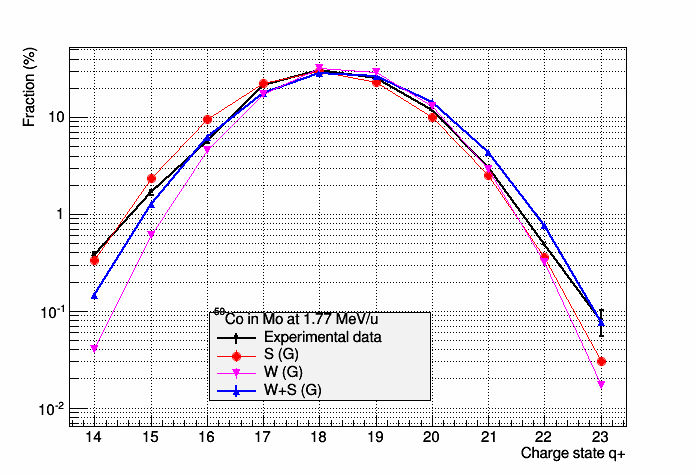}\label{fig:f2}}
\\
\subfloat[]{\includegraphics[width=0.49\textwidth]{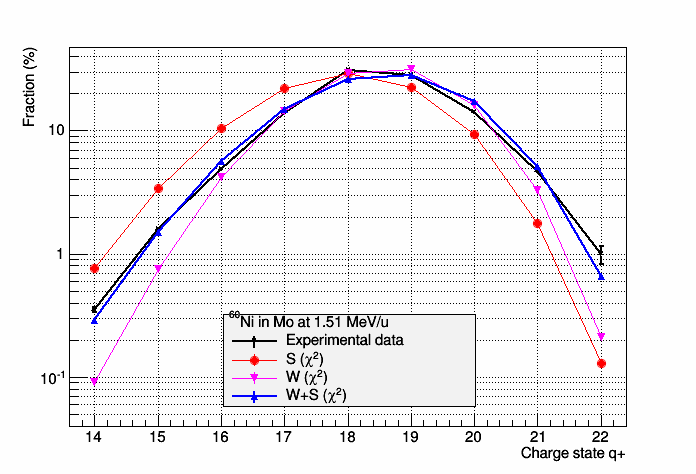}\label{fig:f3}}
\subfloat[]{\includegraphics[width=0.49\textwidth]{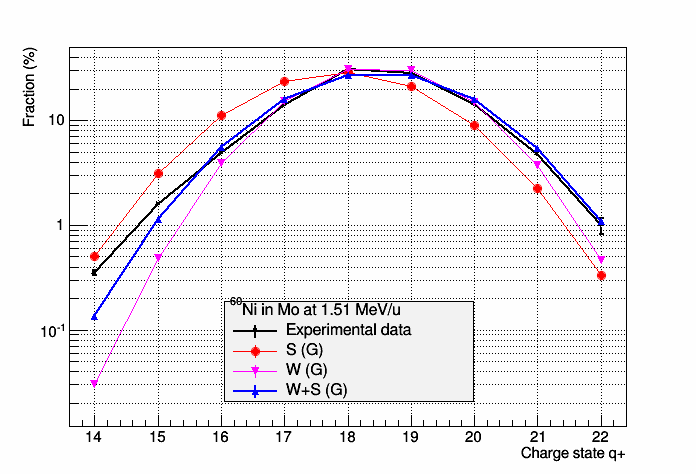}\label{fig:f4}} 
\\
\subfloat[]{\includegraphics[width=0.49\textwidth]{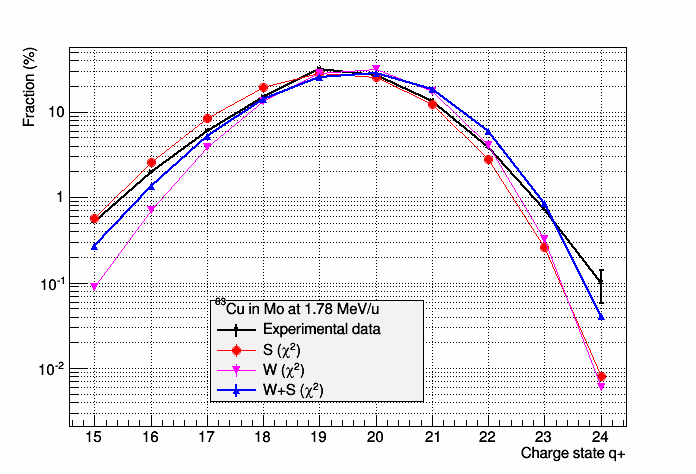}\label{fig:f5}}
\subfloat[]{\includegraphics[width=0.49\textwidth]{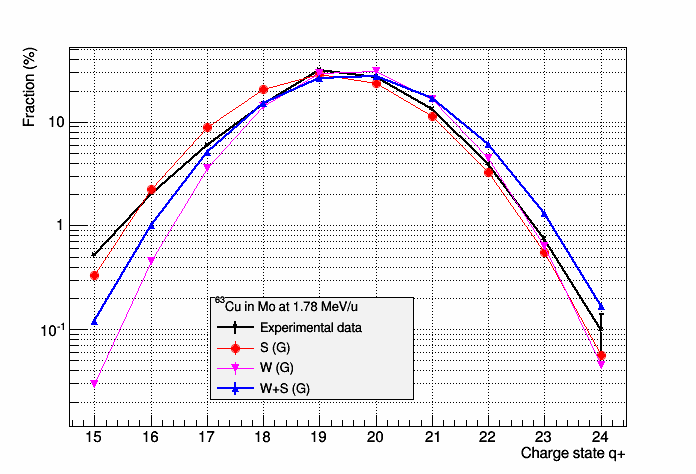}\label{fig:f6}} 
\caption{\label{fig:mult} Calculated CSDs using different semi-empirical models including our ``combinatorial" formulation in comparison with experimental data. The left column of the figure shows calculations assuming a reduced $\chi^{2}$ type shape of the distribution, while the right column shows calculations using a Gaussian shape. Figures \textbf{(a,b)}: Comparison of experimental data from this work to calculations for a $^{59}$Co beam. Figures\textbf{(c,d)}: the same for $^{60}$Ni. Figures \textbf{(e,f)}: The same for $^{63}$Cu. 
In all figures, experimental data are represented by black stars and line, the calculations using the work of Schiwietz et al (S in figure) are presented with red circles and line, the calculations using the work of Winger et al. (W in figure) are presented by magenta inverted triangles and line, and the combinatorial formulation introduced in this work (W+S) is presented with blue upright triangles and line. Values of $\chi^{2}$ for the comparison are presented in table~\ref{tab:table3}.}
    
\end{figure*}

The results of this study support a systematically better agreement with experimental data of the formulation of Schiwietz et al. \citep{sch01} for the distribution widths and of Winger et al. \cite{winger} for the mean charge states $\overline{q}$. Combining these models with a Gaussian or a reduced $\chi^{2}$ distribution function, we explored a more realistic way to reproduce the shape of CSDs in the region of our experimental study (Z$_{p}\sim$28 and Z$_{p}$=42). We compared this ``combinatorial" model to calculations utilizing a single model to estimate both the distribution width \textit{d} and the mean charge state $\overline{q}$. Each of these calculations was performed in two variants; one assuming a Gaussian shaped charge state distribution and one assuming a Baudinet-Robinet type reduced $\chi^{2}$ distribution. The results are presented in Fig. \ref{fig:mult} (a-f) while on Table \ref{tab:table3} the $\tilde{\chi}^2$ values of the calculated CSDs are presented as extracted from chi-square goodness of fit tests. 

From the comparison in Fig. \ref{fig:mult} it is suggested that the combinatorial model combined with a reduced $\chi^{2}$ distribution function produces a qualitatively better description of the experimental data especially in the regions of low intensity charge states near and at the tails of each distribution. At the higher intensity charge states around the mean no significant difference between the reduce $\chi^{2}$ and Gaussian-shape charge distributions is observed. Nevertheless, the overall agreement offered by the combinatorial model (combined with any of the two distribution functions we discussed) is improved in comparison to the other models as demonstrated by the lower on average value of $\tilde{\chi}^{2}$ (see last row of table \ref{tab:table3}).

%%%%%%%%%%%%%%%%%%%%%%%%%%%%%%%%%%%%%%%%%%%%%%%%%%%%%%%%%%%%%%%%%%%%%%%
%%%%%%%%%%%%%%%%%%%%%%%%%%%%%%%%%%%%%%%%%%%%%%%%%%%%%%%%%%%%%%%%%%%%%%%
%%%%%%%%%%%%%%%%%%%%%%%%%%%%%%%%%%%%%%%%%%%%%%%%%%%%%%%%%%%%%%%%%%%%%%%%
%%%%%%%%%%%%%%%%%%%%%%%%%%%%%%%%%%%%%%%%%%%%%%%%%%%%%%%%%%%%%%%%%%%%%%%%

\begin{table*}
\caption{\label{tab:tableA} Fractions percent (\%) of all the measured charge states in the present study. Each row of the table corresponds to a different beam in Mo.}
\begin{center}
\resizebox{\textwidth}{!}{
\begin{tabular}{*{13}{c}}

\hline \hline
 Ion  &E(MeV/u)
 &F$_{14}$&F$_{15}$&F$_{16}$&F$_{17}$&F$_{18}$&F$_{19}$&F$_{20}$
 &F$_{21}$&F$_{22}$&F$_{23}$&F$_{24}$\\ \hline 
 
$^{59}$Co &1.769$\pm$0.005 
&0.38$\pm$0.02&1.67$\pm$0.08&5.71$\pm$0.30 &21.55$\pm$0.37
&30.24$\pm$0.42&24.98$\pm$0.22&11.86$\pm$0.14&3.04$\pm$0.08&0.49$\pm$0.02&0.08$\pm$0.02&-\\   

$^{60}$Ni  &1.510 $\pm$0.009   
 &0.35$\pm$0.02 &1.58$\pm$0.05&4.82$\pm$0.12 &13.97$\pm$0.33&31.04$\pm$0.25
&28.35$\pm$0.51&14.18$\pm$0.21&4.71$\pm$0.16&0.99$\pm$0.17&-&- \\
 
$^{63}$Cu   &1.783 $\pm$0.018  
&-&0.52$\pm$0.02&1.99$\pm$0.05 &5.96$\pm$0.07
&14.78$\pm$0.21&31.74$\pm$0.71&27.06$\pm$0.64&13.23$\pm$0.23&3.88$\pm$0.09
&0.74$\pm$0.02&0.10$\pm$0.04\\

 \hline \hline
% \multicolumn{13}{p{\textwidth}}{}
\end{tabular}
}
\end{center}
\end{table*}
 
%%%%%%%%%%%%%%%%%%%%%%%%%%%%%%%%%%%%%%%%%%%%%%%%%%%%%%%%%%%%%%%%%%%%%
%%%%%%%%%%%%%%%%%%%%%%%%%%%%%%%%%%%%%%%%%%%%%%%%%%%%%%%%%%%%%%%%%%%%%

\begin{table*}
\caption{\label{tab:table1}Comparison of experimentally and theoretically determined values of the mean charge state  $\overline{q}$ in Mo. Predictions of the various semi-empirical models considered in this work are shown. Each row of the table corresponds to a different beam. The last row gives for reference the experimental value from literature for Cu which is in agreement with our own measurement.}

\resizebox{\textwidth}{!}{
\begin{tabular}{*{10}{c}}

\hline \hline
 Ion&E (MeV/u)&Exper.&Schiwietz \citep{sch04}&Shima \citep{shima83}&Baron \cite{baron}&Nik.-Dmit. \citep{nik}&Drouin \cite{drouin}&Winger \citep{winger}&Leon \citep{leon}\\ \hline
 $^{59}$Co&1.769 $\pm$0.005&18.21 $\pm$0.12&18.02&17.73 &17.67&19.12&18.55&18.37&16.59 \\
 $^{60}$Ni&1.510 $\pm$0.009&18.45 $\pm$0.14&17.90&17.57 &17.32&18.96&18.25&18.49&16.26\\
 $^{63}$Cu&1.783 $\pm$0.018&19.34 $\pm$0.20&19.12&18.80 &18.70&20.31&19.66&19.57&17.52\\

 $^{63}$Cu&1.767\footnotemark[1] &19.31\footnotemark[1] \\
 
 \hline \hline
 \multicolumn{10}{p{\textwidth}}{$^{1}$ As reported by K. Shima et al. \citep{shima86}}
\end{tabular}
}

\end{table*}

%%%%%%%%%%%%%%%%%%%%%%%%%%%%%%%%%%%%%%%%%%%%%%%%%%%%%%%%%%%%%%%%%%%%%
\begin{table*}
\caption{\label{tab:table2}Comparison of experimentally and theoretically determined values of the distribution width \textit{d} in Mo. Predictions of the various semi-empirical models considered in this work are shown. Each row of the table corresponds to a different beam. The last row gives for reference the experimental value from literature for Cu which is in agreement with our own measurement. In order to properly compare the predictions of the models with the experimental distribution width, the experimental value of $\overline{q}$ was used as common input in all calculations. 
}
%\footnotesize
\begin{center}
\resizebox{0.85\textwidth}{!}{
\begin{tabular}{cccccccc}
\hline \hline
 Ion&Target&Experimental&Schiwietz \citep{sch01}&Baron \cite{baron}&Nik.-Dmit. \citep{nik}&Winger \citep{winger}&Betz \cite{ben72}\\ \hline
 $^{59}$Co&Mo (foil)&1.31 $\pm$0.02&1.34&1.23 &1.48&1.19&1.40 \\
 $^{60}$Ni&Mo (foil)&1.33 $\pm$0.02&1.37&1.26 &1.52&1.20&1.43\\
 $^{63}$Cu&Mo (foil)&1.35 $\pm$0.02&1.38&1.28 &1.54&1.22&1.45\\
 $^{63}$Cu&Mo (foil)\footnotemark[1] &1.33\footnotemark[1] \\
 \hline \hline \\
\multicolumn{8}{p{0.9\textwidth}}{$^{1}$ As reported by K. Shima et al. \citep{shima86}}
\end{tabular}
}
\end{center}
\end{table*}

%%%%%%%%%%%%%%%%%%%%%%%%%%%%%%%%%%%%%%%%%%%%%%%%%%%%%%%%%%%%%%%%%%%%%%%%
%%%%%%%%%%%%%%%%%%%%%%%%%%%%%%%%%%%%%%%%%%%%%%%%%%%%%%%%%%%%%%%%%%%%%%%% 

\begin{table*}
\caption{\label{tab:table3}$\tilde{\chi}^2$ values of the calculated CSDs. In the calculations were used either combination of models or single models for the $\overline{q}$ and \textit{d}. For explanation of labels see text and figure \ref{fig:mult}:
\textbf{(W)}: Winger et al.,
\textbf{(W+S)}: this work,
\textbf{(S)}: Schiwietz et al.,
\textbf{(G)}: Gaussian distribution,
\textbf{($\chi^{2}$)}: Reduced $\chi^{2}$ distribution.
}
\begin{center}

\resizebox{0.6\textwidth}{!}{

\begin{tabular}{ccccccc}
\hline \hline
&W+S (G)
&W (G)
& S (G)
& W+S ($\chi^{2}$) 
& W ($\chi^{2}$)
& S ($\chi^{2}$)
 \\
\hline

%\hline
$^{59}$Co & 2.48 & 7.22 & 2.58& 3.85&11.31&6.17\\
%\hline
 
$^{60}$Ni & 1.74  & 7.20 & 17.85&1.81&5.92&22.08\\
%\hline

%\hline
$^{63}$Cu & 5.33 & 16.45& 4.12&4.66&9.85&4.93\\
\hline
Average $\tilde{\chi}^2$:&3.18&10.29&8.18&3.44&9.02&11.06\\
\hline \hline

\end{tabular}
}
\end{center}

\end{table*}

%%%%%%%%%%%%%%%%%%%%%%%%%%%%%%%%%%%%%%%%%%%%%%%%%%%%%%%%%%%%%%%%%%%%%%%%
%%%%%%%%%%%%%%%%%%%%%%%%%%%%%%%%%%%%%%%%%%%%%%%%%%%%%%%%%%%%%%%%%%%%%%%%

\section{\label{sec:Conclusion}Conclusion}

Equilibrium charge state distributions of $^{59}$Co, $^{60}$Ni, and $^{63}$Cu beams passing through a 1$\mu$m thick Mo foil
have been measured. A variety of semi-empirical models
for the mean charge state $\overline{q}$ and the distribution width \textit{d}
of equilibrium charge state distributions were compared
with our experimentally determined charge state distributions. Furthermore, our study suggests that an improved agreement of the calculated equilibrium CSDs to the experimental data in the region of study (Z$_{p}\sim$28, Z$_{p}$=42, and E$\backsim$ 2 MeV/u) can be obtained by using a combination of models to describe the equilibrium CSDs. In this 
"combinatorial" description, the formulation of Winger
et al. \cite{winger} is used to calculate the mean charge state
q, the work of Schiwietz et al. \citep{sch01} is followed in the
calculation of the distribution width \textit{d}, and a reduced
$\chi^{2}$ or a Gaussian function can be used to describe the shape of the charge state distribution. Despite the improved agreement of
this "combinatorial" model it is still a phenomenological
prescription with obvious limitations. Realistic first-principle based simulations of equilibrium and non-equilibrium CSDs of heavy ions
would be the ideal way to go forward. Until this is reliably possible, it would be beneficial to extend the current set of experimental data to cover the 19$\lesssim Z_{p},Z_{t} \lesssim$54 region at various energies below 10 MeV/u.

\section*{Acknowledgements}
The authors would like to thank Dr. Oleg Tarasov for sharing details about the code LISE++ \cite{tar2004} and for valuable discussions. The authors also acknowledge support from College of Science and Technology at Central Michigan University. This research was supported by the National Science Foundation (PHY-1419765), Michigan State University and the Facility for Rare Isotope Beams.

%% The Appendices part is started with the command \appendix;
%% appendix sections are then done as normal sections
%% \appendix

%% \section{}
%% \label{}

%% If you have bibdatabase file and want bibtex to generate the
%% bibitems, please use
%%
%%  \bibliographystyle{elsarticle-num} 
%%  \bibliography{<your bibdatabase>}

%% else use the following coding to input the bibitems directly in the
%% TeX file.

\section*{References}

\bibliography{csd}
\bibliographystyle{unsrt}

%\begin{thebibliography}{00}

%% \bibitem{label}
%% Text of bibliographic item

%\bibitem{csd}

%\end{thebibliography}
\end{document}